\documentclass[final,1p,times]{elsarticle} 

\usepackage{amssymb}
\usepackage{amsmath}

\usepackage{booktabs}
\usepackage{threeparttable}
\usepackage[hidelinks]{hyperref}
\usepackage{multirow}

\usepackage[capitalize]{cleveref}
\crefname{section}{Sec.}{Secs.}
\Crefname{section}{Section}{Sections}
\Crefname{table}{Table}{Tables}
\crefname{table}{Tab.}{Tabs.}

\journal{Computers in Biology and Medicine}

\begin{document}

\begin{frontmatter}




\title{MADE-for-ASD: A Multi-Atlas Deep Ensemble Network for Diagnosing Autism Spectrum Disorder} 

\author[a]{Xuehan Liu\fnref{eql-contrib}}
\ead{u7094891@alumni.anu.edu.au}
\author[b,d]{Md Rakibul Hasan\fnref{eql-contrib}}
\ead{Rakibul.Hasan@curtin.edu.au}
\author[b,c]{Tom Gedeon}
\ead{Tom.Gedeon@curtin.edu.au}
\author[a,b]{Md Zakir Hossain\corref{corr}}
\ead{Zakir.Hossain1@curtin.edu.au}

\cortext[corr]{Corresponding author}
\fntext[eql-contrib]{Xuehan Liu and Md Rakibul Hasan contributed equally to this work.}

\affiliation[a]{organization={Australian National University},
            city={Canberra},
            postcode={2601},
            state={ACT},
            country={Australia}
            }
            
\affiliation[b]{organization={Curtin University},
            city={Bentley},
            postcode={6102},
            state={WA},
            country={Australia}
            }
\affiliation[c]{organization={Obuda University},
            city={Budapest},
            postcode={1034},
            country={Hungary}
            }
            
\affiliation[d]{organization={BRAC University},
            city={Dhaka},
            postcode={1212},
            country={Bangladesh }
            }

\begin{abstract}
In response to the global need for efficient early diagnosis of Autism Spectrum Disorder (ASD), this paper bridges the gap between traditional, time-consuming diagnostic methods and potential automated solutions. We propose a multi-atlas deep ensemble network, \textit{MADE-for-ASD}, that integrates multiple atlases of the brain's functional magnetic resonance imaging (fMRI) data through a weighted deep ensemble network. Our approach integrates demographic information into the prediction workflow, which enhances ASD diagnosis performance and offers a more holistic perspective on patient profiling. We experiment with the well-known publicly available ABIDE (Autism Brain Imaging Data Exchange) I dataset, consisting of resting state fMRI data from 17 different laboratories around the globe. Our proposed system achieves 75.20\% accuracy on the entire dataset and 96.40\% on a specific subset -- both surpassing reported ASD diagnosis accuracy in ABIDE I fMRI studies. Specifically, our model improves by 4.4 percentage points over prior works on the same amount of data. The model exhibits a sensitivity of 82.90\% and a specificity of 69.70\% on the entire dataset, and 91.00\% and 99.50\%, respectively, on the specific subset. We leverage the F-score to pinpoint the top 10 ROI in ASD diagnosis, such as \textit{precuneus} and anterior \textit{cingulate/ventromedial}. The proposed system can potentially pave the way for more cost-effective, efficient and scalable strategies in ASD diagnosis. Codes and evaluations are publicly available at \url{https://github.com/hasan-rakibul/MADE-for-ASD}.
\end{abstract}





\end{frontmatter}


\section{Introduction}
Autism Spectrum Disorder (ASD) is a prevalent neurodevelopmental condition characterised by challenges in social and communicative abilities, as well as repetitive and hard-to-control behaviours in daily life \cite{amaral2008neuroanatomy}. ASD often co-occurs with various other conditions, such as intellectual impairment, seizures and anxiety. Autistic individuals display a wide range of characteristics, from mild to severe social and communicative differences, along with restricted and repetitive behaviours and interests \cite{kim2011prevalence}. According to a 2022 report endorsed by the World Health Organisation\footnote{\url{https://www.who.int/news-room/fact-sheets/detail/autism-spectrum-disorders}} \cite{zeidan2022global}, approximately one in 100 children worldwide are autistic, a significant increase from the one in 160 reported in 2012 \cite{elsabbagh2012global}. The economic impact on families of autistic individuals is substantial, making ASD a critical public health concern \cite{ou2015employment}.

Diagnosis of ASD at present leverages a variety of techniques, predominantly behavioural assessments such as the Autism Diagnostic Observation Schedule (ADOS) \cite{lord2000autism} and the Autism Diagnostic Interview-Revised (ADI-R) \cite{lord1994autism}. Both ADOS and ADI-R deliver important insights into an individual's communication, social interactions and behaviour. Nevertheless, these methods primarily depend on observing symptoms, making them largely subjective. The process can be both expensive and lengthy. Additionally, variations in symptom presentations and their severity can introduce additional complications, underscoring the need for a diagnostic method that is more efficient and objective \cite{timimi2019deconstructing}. In this context, functional magnetic resonance imaging (fMRI) of the brain could be leveraged because it provides a non-invasive means of examining brain activity. It has the potential to elucidate the intricate neurological deviations representing ASD, which can be used towards developing automated, objective, efficient and early diagnostic methods \cite{deng2022diagnosing,iidaka2015resting}.

The Autism Brain Imaging Data Exchange (ABIDE) consortium releases resting-state fMRI (rs-fMRI) data with T1 structural brain images and demographic information of autistic individuals and Typical Controls (TCs) \citep{di2014autism}. ABIDE I dataset was collected from 17 international research sites, which makes it diverse and comprehensive for understanding the neurological nuances of ASD. Such heterogeneity in the data can help capture the diverse manifestations of ASD across different populations and geographical locations. ABIDE I data, therefore, can strengthen the generalisability of findings and enhance their clinical relevance and applicability. The preference for using ABIDE I data is proven by many prior works on ASD diagnosis \cite{abraham2017deriving,heinsfeld2018identification,parisot2018disease,deng2022diagnosing,wang2020aimafe,khosla2019ensemble,liu2020attentional,dvornek2018combining,eslami2019asd,almuqhim2021asd,anirudh2019bootstrapping,sherkatghanad2020automated}. This paper uses ABIDE I dataset to classify between ASD and TC, hereinafter referred to as ASD diagnosis.

Machine learning (ML) and deep learning (DL) techniques have been increasingly employed to advance the understanding and diagnosis of various neurodevelopmental disorders, including ASD. ML and DL provide an objective and data-driven approach to diagnosis, thereby reducing the reliance on subjective symptom-based criteria. However, building appropriate models with high-dimensional multi-site neuroimaging data, such as ABIDE I, is challenging because of added dimensionality (i.e., different brain regions may have different cues towards ASD diagnosis) and site diversity (i.e., different data collection sites have different data collection protocols). Including other important cues, such as people's demographic information, can potentially improve the predictive performance but make the system development more challenging while accommodating these data with the primary neuroimaging data. Addressing these challenges, we propose a novel DL-based ASD diagnosis workflow from ABIDE I brain fMRI data and corresponding demographic information. The input rs-fMRI images are processed to extract regions of interest (ROIs) according to three different atlases (brain parcellations), from where functional connectivity features are extracted. This paper's novelty is two-fold. \textbf{(1)} It proposes \textit{MADE-for-ASD}, consisting of a stacked sparse denoising autoencoder (SSDAE) and multi-layer perceptron (MLP), followed by a weighted ensemble learning framework. We incorporate demographic information into the model, which enhances the performance through personalised prediction. \textbf{(2)} This paper provides visualised insights into the most significant ROIs with a high correlation with ASD, which could further assist in understanding the neurobiological underpinnings of ASD.
 
\section{Background and Related Work}
\subsection{Neuroimaging in ASD Diagnosis}
Magnetic resonance imaging (MRI) is a critical tool for understanding the pathophysiology of neurological disorders such as schizophrenia and autism. MRI is valuable due to its cost-effectiveness and non-invasive nature, which have led to its widespread acceptance and application in the medical community \cite{crosson2010functional}. Specifically, functional MRI (fMRI) tracks changes in blood oxygen levels over time, making it adept at inferring brain activity and drawing considerable attention from researchers studying brain dysfunctions \cite{greicius2008resting}. Changes in the intensity of fMRI images throughout the acquisition period usually serve as a representation of brain activity, typically expressed as a time series. Brain disorders rarely manifest as anomalies in singular or multiple brain regions; they typically manifest as atypical connectivity among various brain regions. In this context, functional connectivity helps investigate the association of specific activities between brain regions and has found widespread use in the classification of brain disorders \cite{du2018classification}.

Several datasets are instrumental in advancing autism research by providing extensive neuroimaging data. The ABIDE I dataset aggregates resting-state fMRI data from 17 international sites \citep{di2014autism}. ABIDE II expands on this by including additional subjects and sites, further enhancing its utility for ASD research \citep{di2017enhancing}. Various works in ASD diagnosis have leveraged a variety of machine learning (ML) and deep learning (DL) models with rs-fMRI data from the ABIDE datasets \citep{abraham2017deriving,heinsfeld2018identification,parisot2018disease,deng2022diagnosing,wang2020aimafe,khosla2019ensemble,liu2020attentional,dvornek2018combining,eslami2019asd,almuqhim2021asd,anirudh2019bootstrapping,sherkatghanad2020automated}. For example, \citet{abraham2017deriving} experimented with several ML algorithms, including support vector regression and ridge regression, achieving a classification accuracy of 66.8\%. However, these conventional machine learning models have limited ability to learn complex patterns from raw data and do not take advantage of deep representations. In a separate study, \citet{heinsfeld2018identification} embraced a deep learning technique with an autoencoder (AE) and deep neural network (DNN), reaching 70\% accuracy. \citet{eslami2019asd} leveraged an AE with a single-layer perceptron and achieved 70.3\% accuracy, while \citet{almuqhim2021asd} used an AE, achieving 70.8\% classification accuracy. Although these studies have shown the proof of concept of using AEs in this task, their performance remains limited.

Few studies \citep{kong2019classification,chen2015diagnostic,plitt2015functional} utilised a subset of the ABIDE dataset. For example, \citet{plitt2015functional} employed a random forest classifier to predict ASD on only 179 samples and reported a classification accuracy of 95\%. \citet{chen2015diagnostic} selected 252 high-quality data based on criteria such as head motion, artifacts, and signal dropout and reported an accuracy of 91\% using a random forest classifier.

Some studies \citep{khosla2019ensemble,aghdam2019diagnosis} used both ABIDE I and II datasets. For example, \citet{khosla2019ensemble} trained their Convolutional Neural Network (CNN) architecture with the ABIDE-I data and tested their model's performance on the ABIDE-II dataset. \citet{aghdam2019diagnosis} proposed a `mixture of experts' ensemble approach to diagnosing autistic young children (aged 5–10 years) in three dataset conditions: ABIDE I, ABIDE II, and a combination of them.

The National Database for Autism Research (NDAR) is another significant repository, offering a wide range of data types, including neuroimaging data, collected from various studies and institutions \citep{payakachat2016national}. Using this dataset, \citet{li2018early} proposed a multi-channel CNN model for early ASD diagnosis. Despite the variety of neuroimaging datasets available, the ABIDE I dataset remains the most widely used in ML-based ASD diagnosis \citep{khodatars2021deep}.

\subsection{Selection of Brain Atlases}
Atlas-based parcellation, which divides the entire brain into spatially proximate ROIs, offers several advantages in neuroimaging studies \cite{van2010exploring}: it identifies brain regions with significant connectivity differences between groups, examines brain functional organisation, reduces data dimensionality and improves result interpretability by linking specific brain regions to conditions or phenotypes. The Automated Anatomical Labelling (AAL) atlas, with 116 ROIs, is widely used in ASD diagnosis studies \citep{khosla2019ensemble,khandan2023altered,bazay2024assessing,mahler2023pretraining} due to its ability to provide precise anatomical locations essential for identifying structural abnormalities caused by ASD. The Craddock 200 atlas \citep{craddock2012a}, with 200 ROIs, is based on functional connectivity and is particularly useful for analysing resting-state fMRI data \citep{thomas2020classifying,deng2022diagnosing,bazay2024assessing,mahler2023pretraining}, as it clusters the brain into functionally homogeneous regions.

Recent studies have leveraged multiple atlases to capture diverse patterns related to ASD. For instance, \citet{mahler2023pretraining} proposed a multi-atlas framework for ASD classification using resting-state fMRI data, employing three atlases, including AAL and Craddock 200. Similarly, \citet{deng2022diagnosing} utilised the AAL, Craddock 200 and Craddock 400 atlases. Their experiment indicates that Craddock 200 can outperform Craddock 400 in fMRI-based ASD diagnosis. Hence, we selected the Craddock 200 atlas, hereinafter referred to as the CC atlas.

To further enhance the analysis, we leverage the Eickhoff-Zilles (EZ) atlas with 116 ROIs, which has also been utilised in recent ASD diagnosis studies \cite{subah2023comprehensive,ella2023identifying}. Derived from cytoarchitectonic mapping, the EZ atlas combines anatomical and functional data, offering a nuanced view that bridges structural and functional aspects of brain regions \citep{eickhoff2005a}. By integrating the complementary strengths of AAL, CC and EZ atlases to capture different aspects of brain structure and function, our approach ensures a comprehensive analysis crucial for accurate ASD diagnosis. 

In a typical fMRI-based ASD diagnosis pipeline \citep{deng2022diagnosing,khosla2019ensemble}, atlas data is converted to functional connectivity matrices, which measures the degree of synchronised activity between different brain regions based on the time series of resting-state fMRI brain imaging data. While two of the atlases (AAL and EZ) focus on anatomical locations, they also provide a foundational framework for understanding the structural context within which functional connectivity occurs. Accordingly, functional connectivity of AAL and EZ atlases are utilised in ASD diagnosis literature \citep{khosla2019ensemble,deng2022diagnosing,subah2023comprehensive}.

\section{Method}
\subsection{Task Formulation}
Let us consider $X = \{X_\text{fMRI}, X_\text{demog}\}$, where $X_\text{fMRI}$ and $X_\text{demog}$ are the input fMRI images and demographic/phenotypic information, respectively, and $Y = \{Y_1, Y_2\}$, where $Y_1$ and $Y_2$ are ASD and TC classes, respectively. The goal is to develop a binary classifier $\mathcal{F}$ to predict whether a sample input $x_i \in X$ can be classified as $y_i \in Y$.

\subsection{Overall Framework}

\begin{figure*}[!t]
\includegraphics[width=\textwidth]{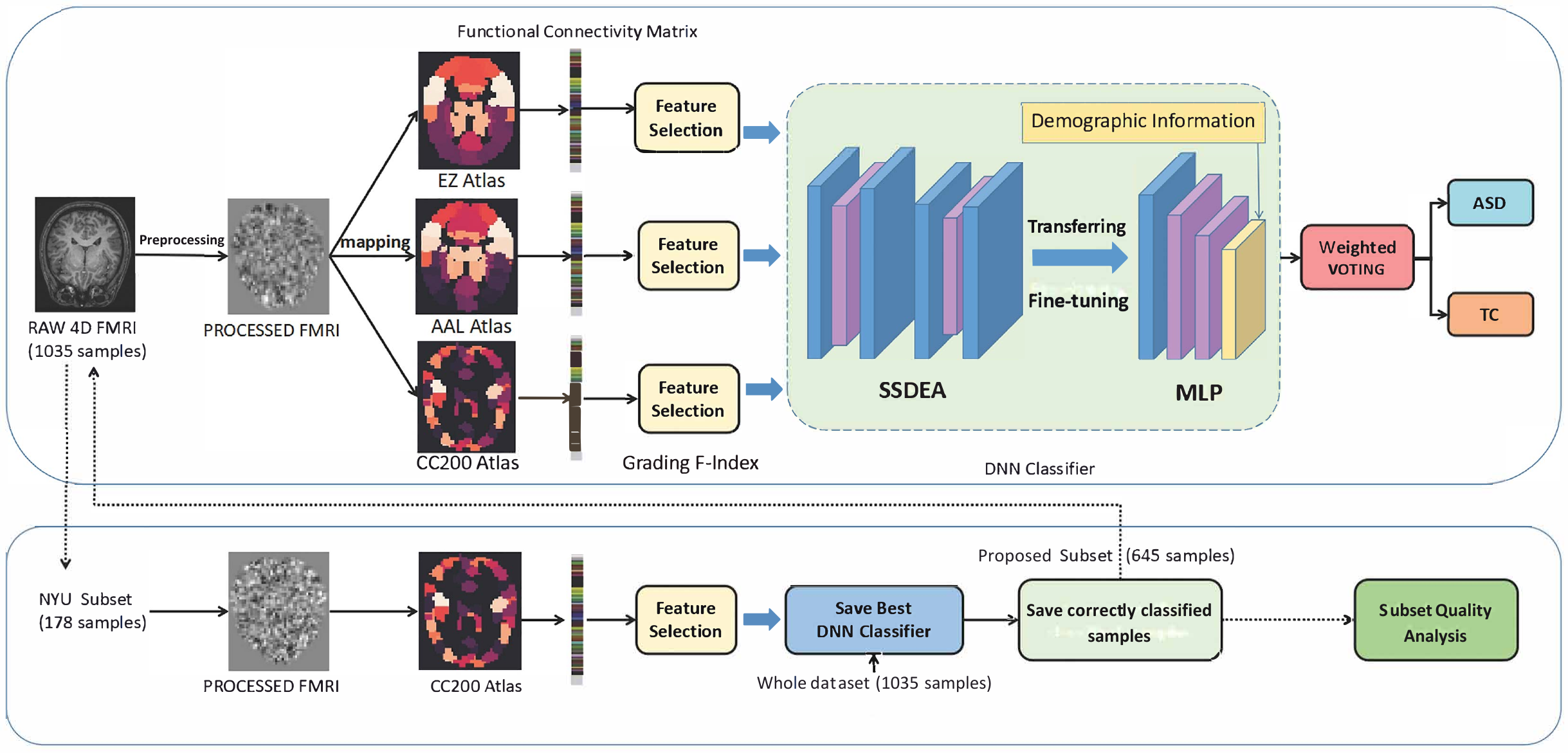}
\caption{Overall framework of our ASD/TC classification workflow. We first calculate functional connectivity matrices on three brain atlases, followed by feature selection using F-score. To diagnose ASD, we use deep learning with a stacked sparse denoising autoencoder (SSDAE) and multi-layer perceptron (MLP), followed by a weighted ensemble. In the case of extracting high-quality data, we train the classifier with the NYU subset and predict on the whole dataset.}
\label{fig:method1}
\end{figure*}

\Cref{fig:method1} illustrates the overall framework of our ASD diagnosis system, which is fundamentally segmented into two distinct stages: the ASD/TC classification phase and the high-quality subset selection phase. 

\subsection{Data Preprocessing}
As data quality control, we exclude all samples with missing fMRI time series. The missing values in the demographic data of the samples are imputed using the mean value of all available data of corresponding categories \cite{donders2006gentle}. In this way, the overall distribution of the data is preserved, and the imputed values are likely to be close to the true values, assuming that the missingness is completely random. While we used the mean value to impute missing data, a more sophisticated approach, such as MICE \citep{azur2011multiple}, could be explored in future work.

We extract the mean time series of ROIs for each sample. We use parcellated regions as our targeted ROIs to extract voxel-level connectivity features. For each atlas, a respective connectivity matrix is formed, which is then condensed into a vector before being inputted into our model. The primary feature we use to differentiate between ASD and TC subjects is functional connectivity. We compute Pearson correlation coefficients between the time series of each pair of brain regions to produce a connectivity matrix:
\begin{equation}
\operatorname{PCC}\left(u, v\right)  =  \frac{E(u v)-E(u) E(v)}{\sqrt{E\left(u^{2}\right)-E^{2}(u)} \sqrt{E\left(v^{2}\right)-E^{2}(v)}}%
\label{eq:data1}
\end{equation}
where $\operatorname{PCC}\left(u, v\right)$ stands for the Pearson correlation coefficient between time series of two brain regions $u$ and $v$, and $E(\cdot)$ is the mathematical expectation. These coefficients range from $-1$ to $1$; coefficients near 1 indicate a strong positive correlation, while those near $-1$ indicate a strong negative correlation between the time series of two brain regions. For example, we obtain a $200\times200$ symmetric matrix for correlation in the case of the CC atlas because it has been divided into $200$ regions. This functional connectivity matrix is used as a feature to classify subjects into ASD and TC groups.

To estimate the duplicated values in the matrix, we take the upper triangle of the symmetric matrix as the original feature representation of this subject. We then flatten the remaining triangle by collapsing it in a one-dimension vector to retrieve a vector of features:
\begin{equation}
S  =  \frac{(N-1) N}{2} \label{eq:data2}
\end{equation}
where $S$ is the dimension for the flattened vector, and $N$ is the number of the regions in the atlas. For example, we get a 1-D vector with $19,900$ features for each sample for the CC atlas. Following the previous computational process, for every individual subject, we secure three functional connectivity feature representations based on the respective three atlases.

\subsection{Feature Selection}

Using F-score \citep{chen2006combining}, we rank all features in descending order to prioritise those with the highest discriminative power between ASD and TC subjects. Mathematically, the F-score is a ratio of variance between groups to variance within groups, quantifying each feature's discriminatory power. A higher F-score indicates a larger difference in means relative to variability, suggesting better class distinction. 
We compute the F-score values for all features in the dataset. Let $x_k$ represent the training vectors, with $k$ ranging from 1 to $m$, $n^+$ being the count of positive instances, and $n^-$ the count of negative instances. The F-score for the $i^{th}$ feature is computed as follows:
\begin{equation}
F(i) = \frac{\left(\bar{r}_{i}^{(+)}-\bar{r}_{i}\right)^{2}+\left(\bar{r}_{i}^{(-)}-\bar{r}_{i}\right)^{2}}{\frac{1}{n_{+}-1} \sum_{k = 1}^{n_{+}}\left(r_{k, i}^{(+)}-\bar{r}_{i}^{(+)}\right)^{2}+\frac{1}{n-1} \sum_{k = 1}^{n-}\left(r_{k, i}^{(-)}-\bar{r}_{i}^{(-)}\right)^{2}}\label{eqn:f_score}
\end{equation}
where $\bar{r}_i$, $\bar{r}_{i}^{(+)}$, and $\bar{r}_{i}^{(-)}$ denote the average of the $i^{th}$ feature for the complete dataset, the positive subset and the negative subset, respectively. The numerator captures the discriminatory power between the positive and negative subsets, while the denominator encapsulates the dispersion within each subset. 

We retain the top 15\% of ranked features. To determine this feature retention range, we conducted a series of tests with selections ranging from 5\% to 50\% based on different parcellations. The optimal parameters were determined based on the classifier's average performance across different parcellations. Details of this experiment are presented in \ref{app-sec:fscore}.

The dimensions of the post-preprocessing data vary across different atlases; for example, features from the CC atlas have a dimensionality of 19,900, while the AAL and EZ atlases have a significantly smaller dimensionality of 6,670. To account for these differences and to maintain the 15\% feature retention, we adopted an adaptive feature selection approach. We retained the top 1,000 ranked features for the AAL and EZ atlases and the top 3,000 ranked features for the CC atlas.


\begin{figure*}[!t]
\centering
\includegraphics[width=1.0\textwidth]{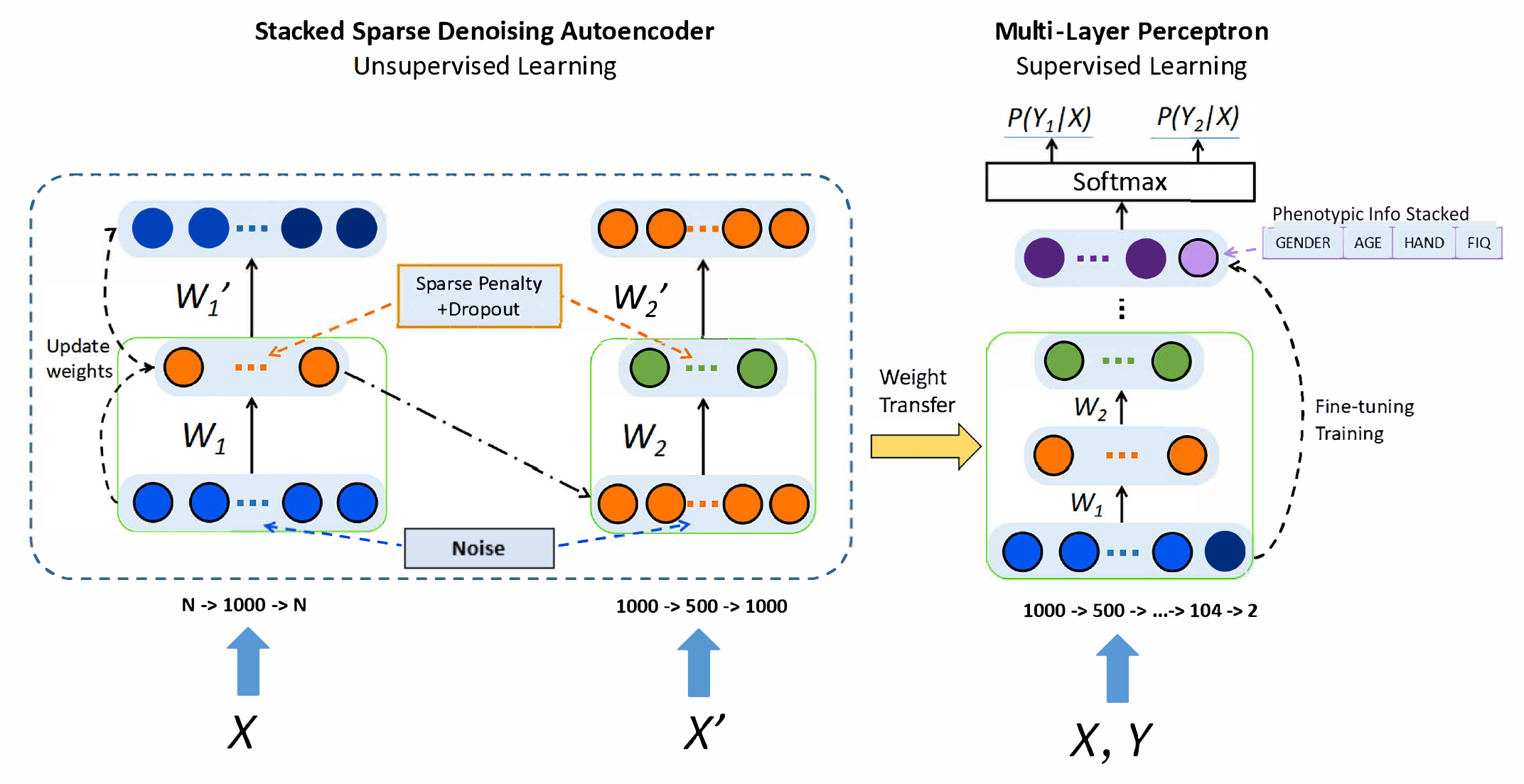}
\caption{Overall architecture and training workflow of the deep networks of the \textit{MADE-for-ASD} model. Knowledge from stacked sparse denoising autoencoders is transferred to a multi-layer perceptron for ASD diagnosis. Here, $X$ refers to the input data, $Y = \{Y_1, Y_2\}$ are ASD and TC classes, and $W$ refers to the weight parameter.}
\label{fig:method3}
\end{figure*}

\begin{table}[!t]
\caption{Parameter configurations of the Autoencoders (AE1, AE2) and Multi-Layer Perceptron (MLP) of the \textit{MADE-for-ASD} model. Here, \textit{`n'} represents the number of input features.}
\centering
\begin{tabular}{@{}lccc@{}} \toprule
\textbf{Parameter} & \textbf{AE1} & \textbf{AE2} & \textbf{MLP} \\ \midrule
Structure & n-1000-n & 1000-600-1000 & n-1000-500-104-2 \\
Hidden layers & 1 & 1 & 3 \\
Learning rate & 0.001 & 0.001 & 0.0005 \\
Batch size & 100 & 10 & 10 \\
Training iteration & 700 & 1000 & 200 \\
Optimiser & GD & GD & SGD \\
Loss function & Cross entropy & Cross entropy & Cross entropy \\
Activation function & tanh & tanh & tanh \\
Momentum range & / & / & 0.1-0.9\\
Dropout rate & 0.5 & 0.5 & 0.3 \\
Sparsity parameter & 0.5 & 0.5 & / \\
Sparse penalty & 0.2 & 0.2 & / \\
Noise proportion & 0.3 & 0.1 & / \\ \bottomrule
\end{tabular}
\label{tab:method1}
\end{table}

\subsection{The MADE-for-ASD Model}
The \textit{MADE-for-ASD} model consists of two primary components, as illustrated in \Cref{fig:method3}. The initial component is an unsupervised learning stage employing a Stacked Sparse Denoising Autoencoder (SSDAE). 
Each layer is trained independently, capturing key variations in the data, with sparsity constraints promoting sparse, distributed representations.

The second component is a supervised learning stage using a Multi-Layer Perceptron (MLP). The parameters learned from the SSDAE are transferred to the MLP's first two layers, followed by fine-tuning to enhance performance on the ASD classification task.

\subsubsection{Stacked Sparse Denoising Autoencoder}
An Autoencoder (AE) consists of input, hidden and output layers, with the hidden layer containing fewer neurons. 
Once feature representation is obtained in an AE, it can be used to train a new AE, leading to a stacked AE with multiple layers.
A Stacked Sparse Denoising Autoencoder (SSDAE) incorporates sparsity constraints and noise addition to the input data for regularisation, enhancing model robustness. The objective function includes a reconstruction loss term and a sparsity penalty term, defined as:
\begin{equation}
J(\Theta) = \frac{1}{m}\sum_{i=1}^{m}L(x^{(i)},\hat{x}^{(i)}) + \beta \sum_{j=1}^{s}\operatorname{KL}(\rho||\hat{\rho}_j)\label{eq:method2}
\end{equation}
where $L(x^{(i)},\hat{x}^{(i)})$ denotes reconstruction loss, $\beta$ is the sparsity weight and $\operatorname{KL}(\rho||\hat{\rho}_j)$ is the Kullback-Leibler divergence between target sparsity $\rho$ and average activation $\hat{\rho}_j$ of hidden unit $j$. The Kullback-Leibler divergence can be computed as:
\begin{equation}
\operatorname{KL}(\rho||\hat{\rho}_j) = \rho \log \frac{\rho}{\hat{\rho}_j} + (1-\rho) \log \frac{1-\rho}{1-\hat{\rho}_j}\label{eqn:kl}
\end{equation}

We employ an SSDAE with two hidden layers for unsupervised pre-training, as shown in the left part of \Cref{fig:method3}. Optimal model performance on the validation set is achieved using reconstruction loss (mean squared error). The input and output layers have $N$ features, where $N$ is the number of input features. The configurations and parameters are detailed in \Cref{tab:method1}.

\subsubsection{Transfer Learning to MLP}
The SSDAE knowledge is transferred to an MLP with three hidden layers. The first two layers, with 1,000 and 500 units, inherit the SSDAE parameters, while the third layer is initialised with random weights. Four demographic features are added to the third layer as additional input. We add these into the last two layers because these features are less significant if we add them into the original input compared to the large amounts of other features. The right section of \Cref{fig:method3} depicts the supervised training stage.

Fine-tuning adjusts the MLP weights to minimise prediction error in supervised tasks. The output layer consists of two units representing the likelihood of ASD or TC, using a \texttt{softmax} activation function to normalise the output distribution and enable the outputs to represent the corresponding probabilities belonging to a particular class. The configurations of the MLP parameters are shown in \Cref{tab:method1}.

\subsubsection{Weighed Ensemble Voting}

\begin{figure}[!t]
\centering
\includegraphics[width=0.7\textwidth]{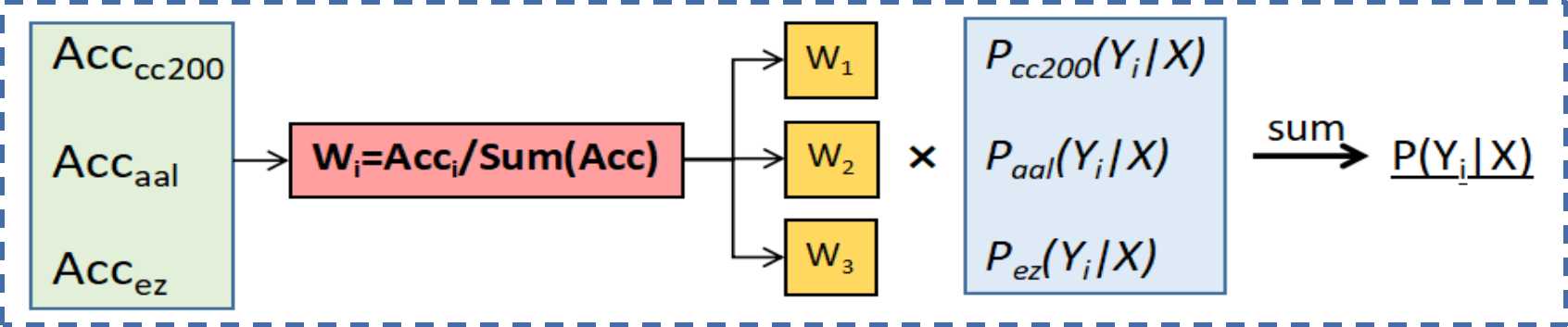}
\caption{The multi-atlas weighted ensemble voting strategy.}
\label{fig:method4_ensemble}
\end{figure}

Ensemble learning uses individual models and solves the same problem.
This work adopts the bagging ensemble approach, which involves creating multiple subsets of the original data, training a model on each and combining their predictions, often through majority voting, to form a final prediction \cite{rokach2010ensemble}. 

Voting can be hard or soft. In hard voting, each classifier in the ensemble votes for one class label, and the class label that gets the majority of votes is predicted. Our method is based on soft voting, as shown in \Cref{fig:method4_ensemble}, which predicts based on the probabilities for each class label. We assign a weight for each classifier according to their individual classification accuracy among all three classifiers:
\begin{equation}
w_{i}=\frac{\text{acc}_{i}}{\sum \text{acc}_{i}} \label{eq:method4}
\end{equation}

We then compute the sum of the products of weights and probabilities corresponding to each class across all classifiers. Subsequently, the class with the highest cumulative value is designated as the output category:
\begin{equation}
y=\operatorname{argmax}_{j \in Y} \sum w_{i}\cdot \mathbb{P}\left(\mathcal{F}_{i}(x)=j\right) \label{eq:method5}
\end{equation}
where $\mathbb{P}\left(\mathcal{F}_{i}(x)=j\right)$ is the predicted probability that instance $x$ belongs to class $j$ according to classifier $\mathcal{F}_{i}$. 


As for the evaluation metrics, we report classification performance in terms of sensitivity and specificity in addition to accuracy because of the unbalanced nature of the dataset.

\section{Experiments}

\subsection{Dataset}
We use rs-fMRI and demographic data from ABIDE I \cite{di2014autism}, the first phase of ABIDE, with 505 autistic individuals and 530 Typical Controls (TCs). We include four key demographic features of the subjects: age (years), sex (male/female), handedness (the dominant hand; left/ambiguous/right) and full-scale IQ (overall intellectual ability) (see \Cref{tab:data1} and \Cref{tab:missing} in \ref{app-sec:data} for the distribution of demographic information across ASD and TC classes and the number of missing demographic data, respectively.). We also employ a subset of the T1w MRI images from the ABIDE I site, NYU Langone Medical Center, to test our proposed methodology. This subset encompasses 182 subjects, including 78 ASD and 104 TC subjects. 
We selected this specific subset because it contains the largest amount of data among all participatory sites in ABIDE I. Additionally, previous studies \citep{kong2019classification,plitt2015functional} have reported ASD classification performance using the NYU subset, which allows us to compare our model's performance with theirs. Refer to \Cref{tab:data2} in \ref{app-sec:data} for details of the NYU subset having 78 autistic individuals and 104 TC subjects.

\subsubsection{Preprocessing}
Preprocessed ABIDE I data with four different pipelines are released under Preprocessed Connectomes Project\footnote{\url{http://preprocessed-connectomes-project.org/abide/}} \cite{craddock2013neuro}. The number of data varies across the resultant datasets from different pipelines. It is important to select a specific pipeline for a fair comparison with the prior works. Most of the prior works \cite{abraham2017deriving,heinsfeld2018identification,sherkatghanad2020automated,eslami2019asd,parisot2018disease,almuqhim2021asd,anirudh2019bootstrapping,liu2020attentional,khosla2019ensemble,wang2020aimafe,deng2022diagnosing} leveraged CPAC (Configurable Pipeline for the Analysis of Connectomes) pipeline. A few studies, however, experimented with other pipelines, such as CCS (Connectome Computation System) by \citet{dvornek2018combining} and DPARSF (Data Processing Assistant for Resting-State fMRI) by \citet{mahler2023pretraining}. The CPAC pipeline includes several operations, such as slice timing correction, motion correction and voxel intensity normalisation. In addition, the nuisance signal was removed utilising 24 motion parameters, CompCor with five components, low-frequency drifts (linear and quadratic trends), and the global signal as regressors. Functional data underwent band-pass filtering ($0.01-0.1$ Hz) and spatial registration using a non-linear method to a template space (MNI152). We, therefore, leverage the CPAC pipeline, hereinafter referred to as the ABIDE I CPAC data. 

Although \citet{di2014autism} released the ABIDE I dataset with 1,112 samples, preprocessing and quality control reduced this number to 1,035, which is consistent with prior studies \citep{heinsfeld2018identification,sherkatghanad2020automated,eslami2019asd,almuqhim2021asd}. For the NYU subset, our preprocessing and quality control reduced the number of samples from 182 to 175.

\subsubsection{Subset Selection}

Previous research \citep{kong2019classification,plitt2015functional} demonstrated better performance with the NYU subset than the whole ABIDE I dataset. This suggests that the NYU subset may possess lower noise levels and higher data quality. Based on this, we hypothesise that using the NYU subset for training could improve the model's ability to select high-quality data from the entire ABIDE I dataset. We, therefore, train our model on the NYU dataset using the CC atlas, save this model and repurpose it as a selector. We apply this pretrained model to classify ASD/TC instances across the whole ABIDE I dataset, which correctly classifies 645 out of 1,035 samples (around 62.3\%). We separate these correctly classified samples to create a new subset, which potentially has higher quality. We refer to it as our \textit{proposed subset} throughout this paper. Analyses on the quality of our proposed subset compared to the whole dataset and the NYU subset are presented in \ref{app-sec:quality}.

While this subset selection aims to improve data quality, it may also lead to the selection of easier examples, potentially inflating performance metrics. Furthermore, this method may introduce exclusion criteria that could affect the generalisability of the results. A thorough analysis of the discarded subset to assess whether it excludes TC, ASD, or both is warranted, which we leave for future work.




\subsection{ASD vs TC Classification}
We evaluate our \textit{MADE-for-ASD} model through experiments on different input data, ablation studies and comparative analyses with state-of-the-art methods. Following the approach of prior ASD diagnosis studies using the ABIDE I dataset \citep{wang2020aimafe,liu2020attentional,anirudh2019bootstrapping,almuqhim2021asd,eslami2019asd,heinsfeld2018identification,parisot2018disease,abraham2017deriving,sherkatghanad2020automated}, we employ 10-fold cross-validation to ensure robust evaluation and mitigate overfitting. This technique splits the dataset into ten equal-sized subsets, with the model being trained on nine subsets and tested on the remaining subset. This process is repeated ten times, and the classification scores are averaged across all runs, providing a cross-validated performance measure.

\subsubsection{Classification Result}
\Cref{tab:res1} reports accuracy, sensitivity and specificity on the whole ABIDE I dataset, our proposed subset and the NYU subset.

\begin{table}[!t]
\caption{Classification performance of \textit{MADE-for-ASD} model in terms of accuracy, sensitivity and specificity on the whole ABIDE I data and its subsets.}
\centering
\resizebox{\linewidth}{!}{%
\begin{tabular}{lcccc} \toprule
\textbf{Data} & \textbf{\# of Subjects} & \textbf{Accuracy (\%)} & \textbf{Sensitivity (\%)} & \textbf{Specificity (\%)}\\ \midrule
Whole dataset & 1,035 & 75.20 & 82.90 & 69.70 \\
Proposed subset & 645 & 88.71 & 81.82 & 92.50 \\
NYU subset & 175 & 96.40 & 91.00 & 99.50 \\ \bottomrule
\end{tabular}%
}
\label{tab:res1}
\end{table}

The accuracy and specificity of the whole dataset are lower than those of other subsets, which can be attributed to the noise and complexity in the whole dataset. The classifier reached its peak performance on the NYU subset, which can be attributed to the high quality of NYU data. This is in line with \citet{kong2019classification} that the classification performance is higher with the NYU subset.


\subsubsection{Comparative Analysis}

\begin{table}[!t]
\caption{Comparison with previous studies on whole ABIDE I CPAC data in terms of classification accuracy.}
\centering
\resizebox{\linewidth}{!}{
\begin{tabular}{llcc} \toprule
\textbf{Study} & \textbf{Model} & \textbf{\# of Subjects} & \textbf{Accuracy (\%)} \\ \midrule
\citet{abraham2017deriving} & SVC & 871 & 66.80 \\
\citet{heinsfeld2018identification} & AE+DNN & 1,035 & 70.00\\
\citet{sherkatghanad2020automated} & CNN & 1,035 & 70.22 \\
\citet{eslami2019asd} & AE+SLP & 1,035 & 70.30 \\
\citet{parisot2018disease} & GCN & 871 & 70.40 \\
\citet{almuqhim2021asd} & SAE+MLP & 1,035 & 70.80\\
\citet{anirudh2019bootstrapping} & Ensemble GCN & 872 & 70.86 \\
\citet{liu2020attentional} & Extra-Trees & 1,054 & 72.20 \\
\citet{wang2020aimafe} & Ensemble MLP & 949 & 74.52 \\
\citet{deng2022diagnosing} & 3D-CNN & 860 & 74.53 \\
\textbf{Ours (MADE-for-ASD)} & \textbf{SSDAE+MLP} & \textbf{1,035} & \textbf{75.20} \\
\bottomrule
\end{tabular}
}
\label{tab:res2}
\end{table}

We compare the performance of our model trained on the whole ABIDE I dataset with the CPAC pipeline, which exhibits encouraging insights (\Cref{tab:res2}). \textbf{(1)} Our model achieved an accuracy score of 75.20\%, which is a boost of 4.4 percentage points over the best prior work \citep{almuqhim2021asd} on 1,035 samples. \textbf{(2)} Even without considering the use of the same amount of input data, our model outperforms the 74.53\% accuracy on 860 samples reported by \citet{deng2022diagnosing}. \textbf{(3)} By incorporating ensemble techniques and introducing sparsity in the AE, our model outperforms other AE and DNN-based approaches \citep{almuqhim2021asd,eslami2019asd,heinsfeld2018identification} by a margin of at least 4.4 percentage points.

We further compare our work with similar studies on certain subsets of the ABIDE I dataset \Cref{tab:res-subset}. Similar to the whole dataset, our proposed \textit{MADE-for-ASD} model achieved the SOTA result on the NYU subset, showing an accuracy of 96.40\%. Besides, our model's performance on the new subset generated by our data selection achieved an accuracy of 88.71\%. This proves our model's ability to perform consistently well on different data subsets.

\begin{table}[!t]
\caption{Comparison with previous studies on a subset of the ABIDE I dataset in terms of classification accuracy.}
\centering
\resizebox{\linewidth}{!}{%
\begin{threeparttable}
\begin{tabular}{llclc} \toprule
\textbf{Study} & \textbf{Data} & \textbf{\# of Subjects} & \textbf{Model} & \textbf{Accuracy (\%)} \\ \midrule
\citet{kong2019classification} & NYU & 182 & SAE & 90.30\\
\citet{chen2015diagnostic} & Selected subset\tnote{a} & 252 & Random Forest & 91.00 \\
\citet{plitt2015functional} & NYU, USM, UCLA & 178 & Random Forest & 95.00  \\
\textbf{Ours (MADE-for-ASD)} & NYU & 175 & \textbf{SSDAE+MLP} & \textbf{96.40} \\
\textbf{Ours (MADE-for-ASD)} & Proposed subset & 645 & \textbf{SSDAE+MLP} & \textbf{88.71} \\ \bottomrule
\end{tabular}
\begin{tablenotes}
    \item[a] Subset was selected based on specific criteria of high-quality data (e.g., low head motion) 
\end{tablenotes}
\end{threeparttable}%
}
\label{tab:res-subset}
\end{table}

This result also underlines the potential influence of data quality on classification outcomes. The performance difference across subsets reinforces the need for data selection in such classification tasks. Our strategy of creating a new subset from the ABIDE I dataset has also proven to be effective. The accuracy on this subset signifies the potential of using selective strategies for data preparation to boost model performance.


\subsubsection{Ablation Study}

As can be seen on \Cref{tab:res-ablation}, we carry out a series of ablation experiments to better understand the individual contributions of different components of the \textit{MADE-for-ASD} model. These experiments include removing the ensemble voting and using single and combination of atlas data for training, abolishing the F-score-based feature selection, and omitting the sample demographic information during classification.

\begin{table}[!t]
\caption{Classification accuracy in ablation study by removing atlases, feature selection and demographic information. Negative sign $(-)$ refers to the removal of the corresponding component.}
\centering
\begin{tabular}{@{}lc@{}} \toprule
\textbf{Component} & \textbf{Accuracy (\%)}\\ \midrule
CC + AAL + EZ + Feature Selection + Demographic & 75.20 \\ \midrule
$-$ \{AAL, EZ\} & 73.42\\
$-$ \{CC, EZ\} & 71.20\\
$-$ \{CC, AAL\} & 68.74\\
$-$ EZ & 74.10 \\
$-$ Feature Selection & 72.80 \\
$-$ Demographic & 73.50 \\ \bottomrule
\end{tabular}
\label{tab:res-ablation}
\end{table}

When we remove the voting and use only a single atlas and a combination of atlases, the classification accuracy decreases. When removing data of the AAL and EZ atlases, the accuracy drops to 73.42\%. It drops further to 71.20\% after removing the CC and EZ atlases. Removing the CC and AAL atlases, the accuracy reaches the lowest at 68.74\%. When we only remove the EZ atlas, the accuracy is somewhat increased. This shows that integrating multiple atlases via voting contributes to the improved performance of our model. The different atlas information can complement each other, thereby enhancing the robustness of the classification task.

When we abolish the feature selection based on the F-score, it leads to a decrease in accuracy to 72.80\%. The drop in accuracy demonstrates the significance of feature selection. By focusing on the most informative features and deleting the noisy features, our model can better discriminate between classes. Lastly, removing the demographic information leads to an accuracy of 73.50\%. This indicates that including demographic information can aid in the classification process, further improving the decision boundaries.

Although the inclusion of demographic data leads to better performance, these data should be carefully utilised with ML-based practical ASD diagnosis systems so that the system is not biased to particular demographics. Future work should focus on quantifying and mitigating such biases in ML models predicting ASD. Furthermore, exploring demographic differences within the pipeline -- such as sex differences in the top ROIs revealed by our algorithm -- represents an important avenue for future research. Similarly, evaluating performance on stratified demographic subsets would help ensure that our model performs consistently across different groups.

While our fMRI-based approach with deep learning provides robust diagnostic performance for ASD, it is important to acknowledge the interpretability challenges associated with these methods. Traditional behaviour-based approaches offer specific insights into the symptomatic nuances of participants, which can enhance the substance of clinical reports. In contrast, our method, despite its diagnostic accuracy, may offer fewer insights into individual symptomatic specificities. This limitation is two-fold: first, there is a need for further research to improve the interpretation of fMRI data and its correlation with autism symptoms; second, the deep learning model used in our study offers limited transparency, making it challenging to explain individual predictions based on input features. Future work should enhance fMRI interpretability for autism to bridge the gap between diagnostic accuracy and clinical insight.

\subsection{Visualisation of Top ROIs}

\begin{figure}[!t]
\centering
\includegraphics[width=1\columnwidth]{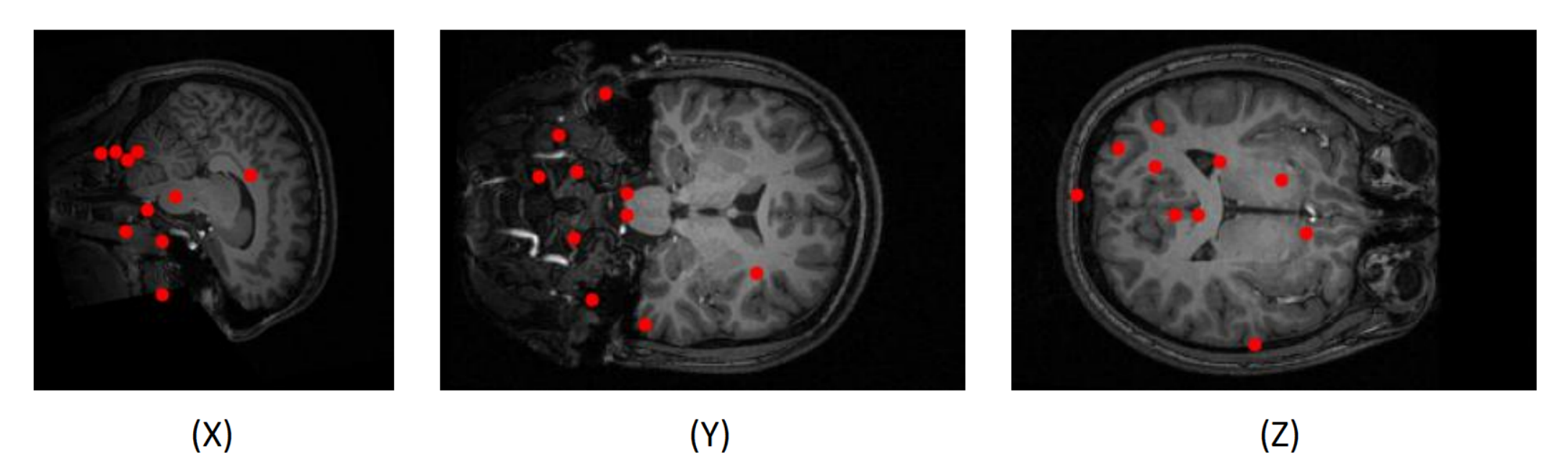}
\caption{Top 10 most significant regions of interest towards ASD diagnosis using fMRI CC200 atlas.}
\label{fig:res1}
\end{figure}

We utilise the F-score feature selection methodology to discern and rank the most influential features.
The initial 3D images from the CC atlas are clustered based on the spatial position of ROIs. These resulting cluster centres are treated as the spatial coordinates of the ROIs. On an existing three-dimensional brain fMRI image from the dataset, we then plot the spatial coordinates of these cluster centres.
We rank the appearance frequencies of the associated ROIs, thereby identifying the top 10 most frequently occurring areas of interest (\Cref{fig:res1}).

As can be seen, the \textit{precuneus}, typically recognised as a central node of the default mode network (DMN) \citep{utevsky2014precuneus}, has a crucial impact on ASD classification. This region is involved in self-referential thought and social cognition, which are often disrupted in autistic individuals \citep{cherkassky2006functional}. Several studies have suggested that DMN connectivity can be associated with a neurophenotype of ASD \citep{khosla2019ensemble}. For example, \citet{chen2015diagnostic} highlighted substantial contributions from default mode and somatosensory areas toward ASD diagnosis. Similarly, \citet{abraham2017deriving} found distinguishing connections in the DMN related to ASD diagnosis using the ABIDE dataset.

The anterior \textit{cingulate/ventromedial} prefrontal cortex, a region with established connections to autism, was notably pronounced in the ASD classification problem \cite{watanabe2012diminished}. This area is associated with emotional regulation and decision-making, processes that are often impaired in ASD \cite{di2014autism}. Furthermore, anomalies in the medial prefrontal cortex node of the DMN have been shown to detect social deficits in autistic children \citep{menon2013developmental}. Additionally, the left parietal cortex was stressed for ASD prediction, aligning with the lateralised activation seen in this region in autistic individuals \cite{koshino2005functional}. These findings from the visualisation of top ROIs corroborate previous studies that highlight the importance of these regions in ASD pathology \cite{khosla2019ensemble}.

\section{Conclusion}

ASD is a neurodevelopmental disorder characterised by a spectrum of symptoms and impairments. Common features of ASD include challenges with social interaction and communication, alongside a preference for repetitive behaviours and interests, highlighting the diverse nature of this condition. This paper proposes a novel ASD diagnosis framework, \textit{MADE-for-ASD}, involving a weighted ensemble of DNNs using multi-atlas brain fMRI data. Through the F-score-based feature selection method, we obtain discriminative features that offer valuable visual insights into significant ROIs associated with ASD. They shed light on the interplay of different features and their respective contribution towards ASD diagnosis. This would help clinicians and researchers gain a more intuitive understanding of how different brain regions contribute to ASD. Our model consists of an SSDAE and an MLP, complemented by integrating demographic information, which significantly enhances our model's predictive capabilities. Furthermore, our method of selecting high-quality classification subsets serves to reduce dataset noise and improve data quality. Our model achieves the SOTA accuracy on both the whole ABIDE I dataset (an improvement of 4.4 percentage points) and its subset. Such an imaging-based ASD prediction system can benefit patients, families and healthcare systems worldwide through objective, efficient, non-intrusive and early diagnosis.


\appendix
\section{Feature Selection using F-score}\label{app-sec:fscore}

\begin{figure}[!ht]
\centering
\includegraphics[width=0.6\textwidth]{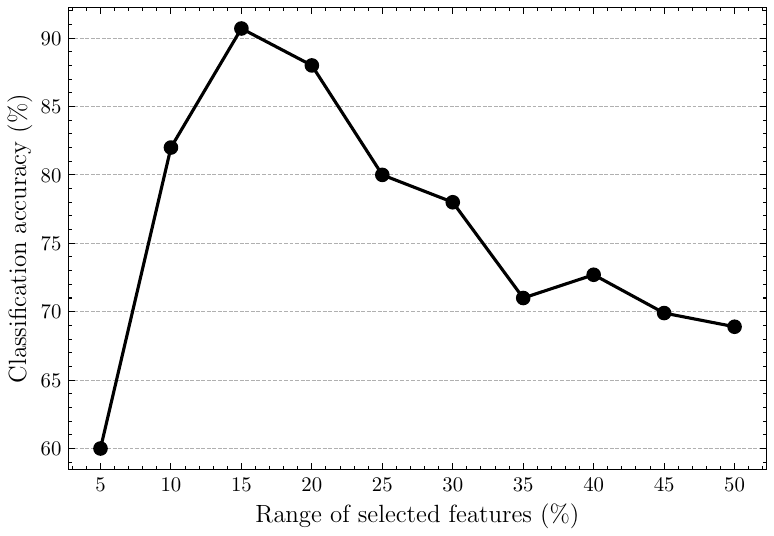}
\caption{ASD classification accuracy with different feature selection ranges in our F-score-based feature selection. The presented accuracy refers to the average accuracy over all three parcellations.}
\label{fig:method2}
\end{figure}

\Cref{fig:method2} illustrates our feature selection process based on F-score, showing that the classification accuracy was highest when using the top 15\% of the total features. A histogram of F-scores for each atlas can reveal the decay and range of feature values, which should be investigated in future work.

\section{ABIDE I Dataset Statistics}\label{app-sec:data}
The demographic details and sample distribution of the ABIDE I whole dataset and the NYU subset are presented in \Cref{tab:data1} and \Cref{tab:data2}, respectively.

\begin{table*}[!ht]
\centering
\caption{Distribution of demographic information of the ABIDE I subjects.}
\label{tab:data1}
\resizebox{\linewidth}{!}{
\begin{threeparttable}
\begin{tabular}{l|*4c|*4c}
\toprule
\multirow{2}{*}{Site} &         ASD &   &         &         &           TC &    &       &    \\ \cline{2-9}
      & Age &   Sex & Hand &  FIQ   &  Age  &  Sex  & Hand & FIQ   \\ \midrule
CALTECH &  27.4 (10.3) &       15/4 &            0/5/14 &         108.2 (12.2) &   28.0 (10.9) &       14/4 &            1/3/14 &  114.8 (9.3) \\
CMU &   26.4 (5.8) &       11/3 &            1/1/12 &         114.5 (11.2) &    26.8 (5.7) &       10/3 &            0/1/12 &  114.6 (9.3) \\
KKI &   10.0 (1.4) &       16/4 &            1/3/16 &          97.9 (17.1) &    10.0 (1.2) &       20/8 &            1/3/24 &  112.1 (9.2) \\
LEUVEN &   17.8 (5.0) &       26/3 &            3/0/26 &         109.4 (12.6) &    18.2 (5.1) &       29/5 &            4/0/30 & 114.8 (12.4) \\
MAX$\_$MUN &  26.1 (14.9) &       21/3 &            2/0/22 &         109.9 (14.2) &    24.6 (8.8) &       27/1 &            0/0/28 &  111.8 (9.1) \\
NYU &   14.7 (7.1) &      65/10 &               NaN &         107.1 (16.3) &    15.7 (6.2) &      74/26 &               NaN & 113.0 (13.3) \\
OHSU & 11.4 (2.2) & 12/0 & 1/0/11 & 106.0 (20.1) & 10.1 (1.1) & 14/0 & 0/0/14 & 115.0 (10.7)\\
OLIN &   16.5 (3.4) &       16/3 &            4/0/15 &         112.6 (17.8) &    16.7 (3.6) &       13/2 &            2/0/13 & 113.9 (16.0) \\
PITT &   19.0 (7.3) &       25/4 &            3/1/25 &         110.2 (14.3) &    18.9 (6.6) &       23/4 &            1/1/25 &  110.1 (9.2) \\
SBL &  35.0 (10.4) &       15/0 &               NaN &                 NaN &    33.7 (6.6) &       15/0 &               NaN &         NaN \\
SDSU &   14.7 (1.8) &       13/1 &            1/0/13 &         111.4 (17.4) &    14.2 (1.9) &       16/6 &            3/0/19 & 108.1 (10.3) \\
STANFORD &   10.0 (1.6) &       15/4 &            3/1/15 &         110.7 (15.7) &    10.0 (1.6) &       16/4 &            0/2/18 & 112.1 (15.0) \\
TRINITY &   16.8 (3.2) &       22/0 &            0/0/22 &         108.9 (15.2) &    17.1 (3.8) &       25/0 &            0/0/25 &  112.5 (9.2) \\
UCLA &   13.0 (2.5) &       48/6 &            6/0/48 &         100.4 (13.4) &    13.0 (1.9) &       38/6 &            4/0/40 & 106.4 (11.1) \\
UM &   13.2 (2.4) &       57/9 &            7/8/51 &         105.5 (17.1) &    14.8 (3.6) &      56/18 &            9/2/63 &  108.2 (9.7) \\
USM &   23.5 (8.3) &       46/0 &               NaN &          99.7 (16.4) &    21.3 (8.4) &       25/0 &               NaN & 115.4 (14.8) \\
YALE &   12.7 (3.0) &       20/8 &            5/0/23 &          94.6 (21.2) &    12.7 (2.8) &       20/8 &            4/0/24 & 105.0 (17.1) \\
\bottomrule
\end{tabular}
\begin{tablenotes}
      \item Age: Average (Standard Deviation), Sex: Male/Female, FIQ: Average (Standard Deviation)
      \item Hand: Left/Ambiguous/Right; the dominant hand of samples 
    \end{tablenotes}
\end{threeparttable}
}
\end{table*}

\begin{table}[!ht]
    \caption{Number of missing demographic data in the ABIDE I dataset.}
    \label{tab:missing}
    \centering
    \begin{tabular}{lc} \toprule
    Demographic information & \# of missing values \\ \midrule
    Age & 0 \\
    Sex & 0 \\
    Handedness & 326 \\
    Full-scale IQ & 72 \\ \bottomrule
    \end{tabular}
\end{table}

\begin{table}[!ht]
\caption{Class-wise statistics of the NYU subset.}
\centering
\begin{tabular}{cccc} \toprule
Class & Size & Age & Gender (F/M) \\ \midrule
ASD & 78 & $14.54 \pm 5.29$ & 10/68 \\
TC & 104 & $15.87 \pm 5.04$ & 26/78 \\
\bottomrule
\end{tabular}
\label{tab:data2}
\end{table}

\section{Subset Quality Analysis}\label{app-sec:quality}

\begin{figure*}[!ht]
\centering
\includegraphics[width=1\textwidth]{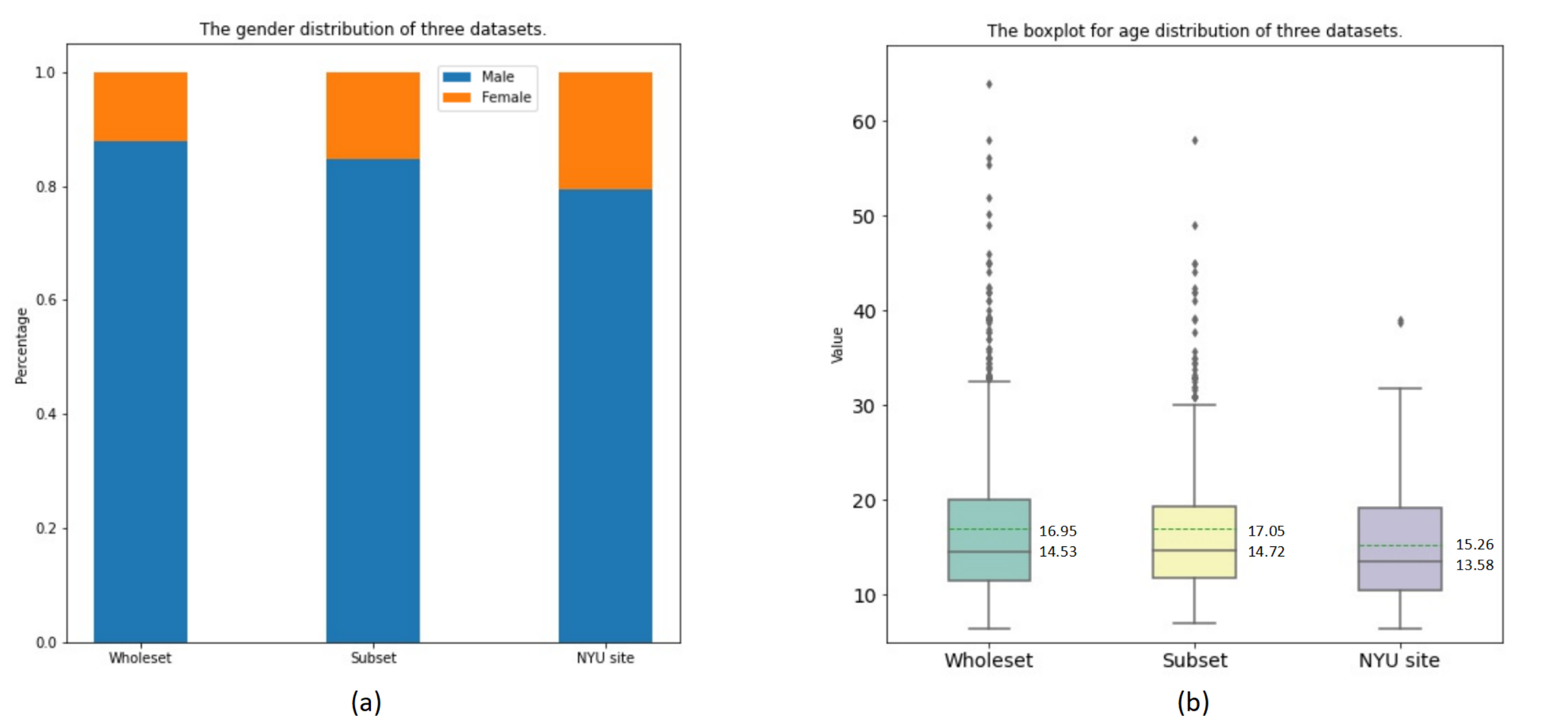}
\caption{Demographic information distribution on different ranges of datasets 
\label{fig:res2}}
\end{figure*}

We analyse the age distribution in the NYU subset, our selected subset and the whole dataset. As can be seen in \Cref{fig:res2}, samples from younger individuals manifest a superior edge in the classification tasks. According to the higher classification accuracy on the NYU subset, the nature of these results suggests that age could play a pivotal role in shaping the outcomes of classification tasks. Given these findings, further exploration into the impact of age distribution on ASD classification tasks may be discussed. 

\begin{figure*}[!ht]
\centering
\includegraphics[width=1\textwidth]{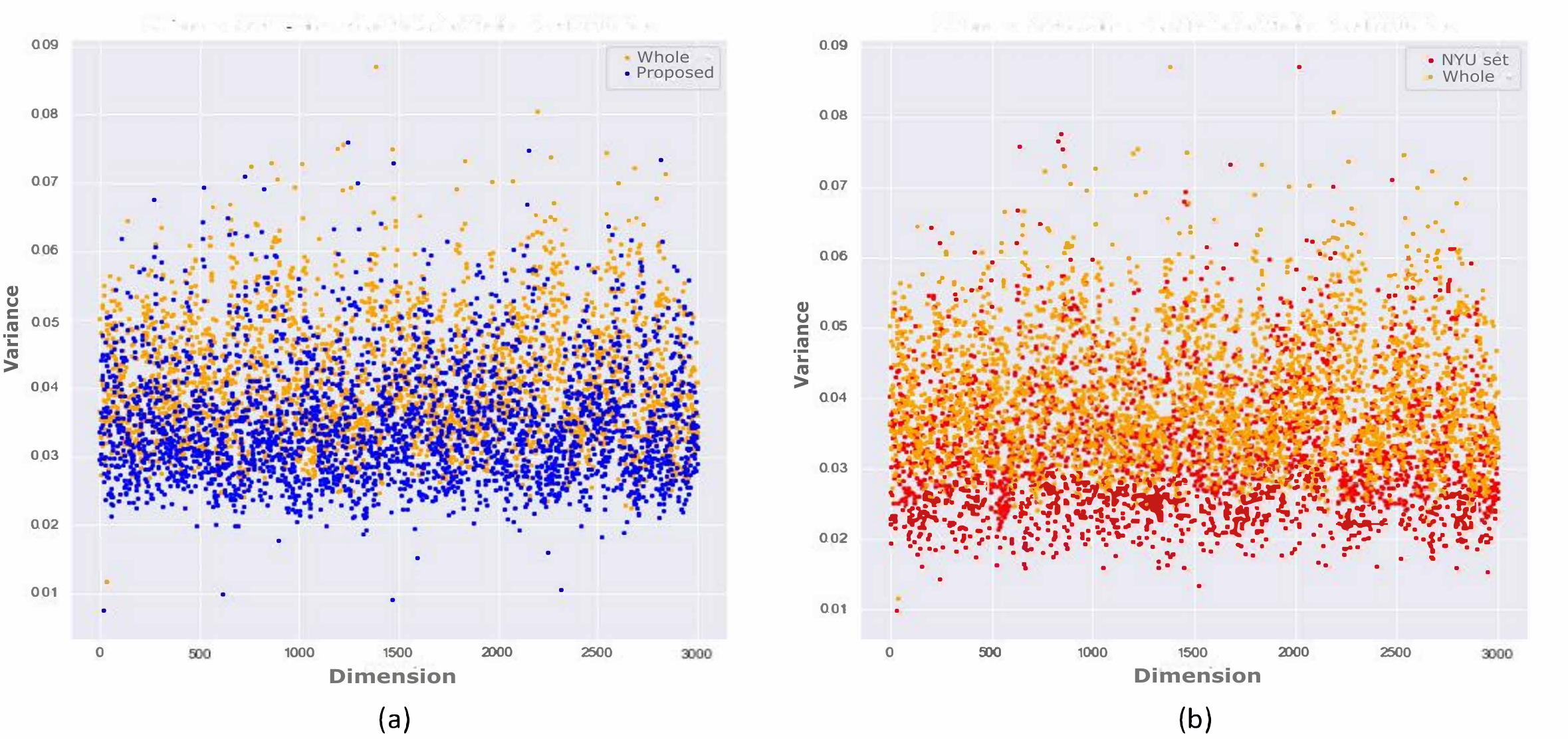}
\caption{Comparison of features variance between the whole dataset and (a) our proposed subset and (b) NYU subset.}
\label{fig:res3}
\end{figure*}

Our selected subset demonstrates a reduction in noise and variance compared to the whole dataset (\Cref{fig:res3}). This reduction potentially suggests higher-quality data because lower variance can enable the model to learn patterns more effectively with more accurate predictions. This aligns with findings that reduced data variability can enhance model performance \cite{domingos2012few}.

\clearpage 

\bibliographystyle{elsarticle-num-names} 
\bibliography{ref.bib}

\end{document}